\documentclass[prl,letterpaper,aps,10pt,superscriptaddress,twocolumn,floatfix,showpacs]{revtex4-1}
\usepackage{graphicx}
\usepackage{amsmath}
\usepackage{amsfonts}
\usepackage{amssymb}
\usepackage{epsfig}
\usepackage[pdftex]{color}
\usepackage{amsmath,graphicx,amssymb,braket,xcolor,subfigure,upgreek}
\usepackage[colorlinks, linkcolor=blue, citecolor=blue, urlcolor=blue, breaklinks=true]{hyperref}
\usepackage{microtype}
\usepackage{bbm}
\usepackage{color}
\newcommand{\fref}[1]{Fig.~\ref{#1}}
\renewcommand{\section}[1]{\emph{\textbf{#1}}. --}

\renewcommand{\max}[0]{\mathrm{max}}

\renewcommand{\emph}[1]{{\it #1}}
\renewcommand{\vec}[1]{\boldsymbol{#1}}

\renewcommand{\vec}[1]{\boldsymbol{#1}}

\newcommand{\beq}{\begin{equation}}
\newcommand{\eeq}{\end{equation}}
\newcommand{\rb}{\vec{r}}

\def\db{\boldsymbol{\wp}}  
\def\dbu{\hat{\boldsymbol{\wp}}} 
\def\rbu{\hat{\bf{r}}}

\def\seg{\sigma^{eg}}
\def\sge{\sigma^{ge}}

\def\Eb{\boldsymbol{E}}

\def\Gb{{\boldsymbol G}}

\begin{document}
\title{Extraordinary subradiance with lossless excitation transfer in dipole-coupled nano-rings of quantum emitters}
\author{Maria Moreno-Cardoner}
\thanks{These authors contributed equally to this work.}
\affiliation{ICFO-Institut de Ciencies Fotoniques, The Barcelona Institute of Science and Technology, 08860 Castelldefels (Barcelona), Spain}
\affiliation{F\'isica Te\`orica: Informaci\'o i Fen\`omens Qu\`antics. Departament de F\'isica, Universitat Aut\`onoma de Barcelona, 08193 Bellaterra, Spain}
\author{David Plankensteiner}
\thanks{These authors contributed equally to this work.}
\affiliation{Institut f\"ur Theoretische Physik, Universit\"at Innsbruck, Technikerstr. 21a, A-6020 Innsbruck, Austria}
\author{Laurin Ostermann}
\affiliation{Institut f\"ur Theoretische Physik, Universit\"at Innsbruck, Technikerstr. 21a, A-6020 Innsbruck, Austria}
\author{Darrick Chang}
\affiliation{ICFO-Institut de Ciencies Fotoniques, The Barcelona Institute of Science and Technology, 08860 Castelldefels (Barcelona), Spain}
\affiliation{ICREA-Instituci\'o Catalana de Recerca i Estudis Avancats, 08015 Barcelona, Spain}
\author{Helmut Ritsch}
\affiliation{Institut f\"ur Theoretische Physik, Universit\"at Innsbruck, Technikerstr. 21a, A-6020 Innsbruck, Austria}
\date{\today}

\begin{abstract}
A ring of sub-wavelength spaced dipole-coupled quantum emitters possesses only few radiant but many extraordinarily subradiant collective modes. These exhibit a 3D-confined spatial radiation field pattern forming a nano-scale high-Q optical resonator. We show that tailoring the geometry, orientation and distance between two such rings allows for increasing the ratio of coherent ring-to-ring coupling versus free-space emission by several orders of magnitude. In particular we find that subradiant excitations, when delocalized over several ring sites, are effectively transported between the rings with a high fidelity. 
\end{abstract}

\pacs{42.50.Ar, 42.50.Lc, 42.72.-g}

\maketitle

\section{Introduction}
An efficient and controllable energy transfer is significant for a wide variety of applications, ranging from solar energy conversion, near-field communication protocols, and photosynthetic processes, to quantum communication, just to name a few. In all these processes it is crucial to find mechanisms that allow for a minimization of the energy loss. 
A concrete example is the coupling between two or more conventional optical ring resonators, which can be used for the realization of switches, high-order optical filters or mechanical sensors. This coupling relies on the evanescent field extending outside the waveguide modes, and it can be enhanced by reducing the distance between them, increasing the coupling length or modifying their refractive indices. In a very different context, in light-harvesting complexes such as LHC-II, an intriguing question is whether nature chooses a particular configuration to minimize the energy loss in the transport of the absorbed photon. 

In this work, we will focus on excitation transport between two independent chains formed by regularly arranged atom-like emitters and show how subradiance can aid this process. In particular, we will show that chains of atoms forming two rings can reduce the energy loss and enhance the fidelity in the excitation transport between them. 

Spontaneous photon emission from a single excited atom in free space is strongly modified by identical emitters in the ground state nearby. Dipole-dipole coupling leads to the appearance of fast decay via superradiant states as well as long-lived subradiant behavior~\cite{lehmberg1970radiation, haroche1982superradiance, zoubi2008bright,asenjo2017exponential,jenkins2016many}. Energy  shifts originating from coherent dipole-dipole coupling~\cite{scully2009collective} lead to a broadening of the collective energy spectrum strongly growing with smaller atomic distances~\cite{plankensteiner2015selective}. The properties of these collective states strongly depend on the geometry of the dipole arrays~\cite{lehmberg1970radiation,zoubi2012optical,ostermann2012cascaded}. While superradiance is a well-known experimental phenomenon, subradiance is much more elusive~\cite{guerin2016subradiance}.

As shown before, excitons become perfectly dark in an infinite chain of sub-wavelength spaced dipoles~\cite{zoubi2010metastability} when their wave-vector surpasses the free-space photon wave-vector. For a finite chain of emitters the maximal excited state lifetime grows with the third power of the atom number N as photon emission only occurs at its ends~\cite{plankensteiner2015selective, asenjo2017exponential}. As excitations are efficiently guided without dissipation as in a thin optical fiber~\cite{zoubi2010hybrid,ostermann2018super} this should have useful applications for efficient optical photon storage~\cite{asenjo2017exponential}. Low loss guiding studies have also been performed for chains of gold nano-particles~\cite{brongersma2000electromagnetic}. 

\begin{figure}[t]
\centering
\includegraphics[width=\columnwidth]{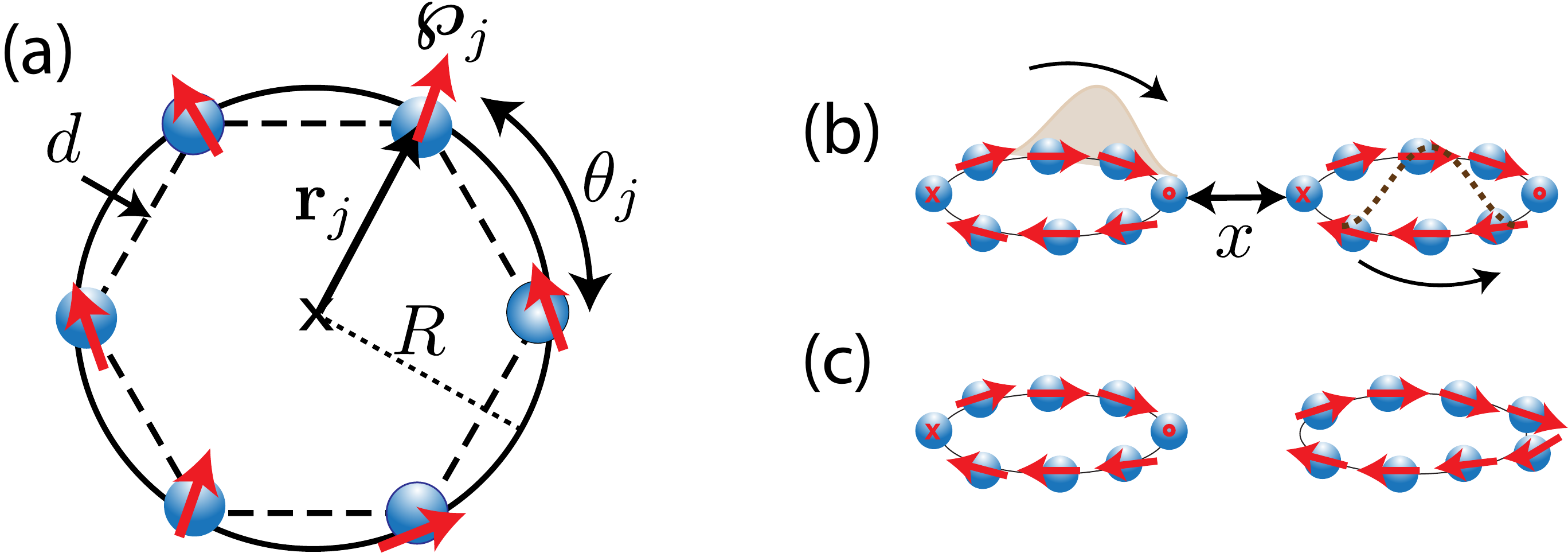}
\caption{\textit{Schematics of the system setup.} {\bf (a)} A single ring with inter-particle distance $d$ and radius $R$. The red arrows denote arbitrary dipole orientations. {\bf(b)} and {\bf (c)} A single excitation is transferred between two in-plane rings separated by the distance $x$. {\bf (b)} and {\bf (c)} correspond to the site-site and site-edge configurations, respectively.}
\label{fig1}
\end{figure}
As a central new phenomenon, we study the excitation transfer between two separate dipole arrays. Although the strong transverse field confinement of subradiant states leads to an exponential suppression with distance between two parallel chains, energy eventually is still transferred between them with an almost negligible loss~\cite{zoubi2010hybrid}. Here, we focus on chains folded into rings, which as shown and discussed in Ref.~\cite{asenjo2017exponential} exhibit extraordinarily long exciton lifetimes with very special radiative properties. A ring of dipoles at small distances in this sense implements a minimalist form of an optical ring resonator as depicted in \fref{fig1}, which, in principle, can exchange energy with a second nearby ring.

As with conventional fiber-optical ring resonators we study how two such rings can be coupled via their mode overlap at a minimal free-space radiation loss. Efficient coherent coupling between two long-lived states is one of the central ingredients needed in distributed quantum computing~\cite{serafini2006distributed}. Surprisingly, the subradiant states of individual rings feature a slower but much more efficient ring-to-ring energy transfer than superradiant states.

Note, that many light-harvesting complexes such as LHC-II in biological systems exhibit a structure made out of coupled dipole rings~\cite{blankenship2014molecular,nelson2005complex,cogdell2006architecture}. While modeling them realistically certainly requires a much more detailed and sophisticated description, a corresponding simplistic model of eight outer rings commonly coupled to an inner ring~\cite{Engel2007EvidenceFW} shows a wealth of complex and nontrivial dynamics with evidence of coherent excitation propagation already~\cite{Engel2007EvidenceFW, Panitchayangkoon2011DirectEO}. In our toy model inspired by this geometry, dark states play an essential role in the coupled dynamics and energy transfer between the rings.

\section{System}
Let us consider $N$ identical two-level quantum emitters with given dipole orientations (denoted by $\dbu_i$, $i=1,\dots, N$) positioned in a regular ring with inter-particle distance $d$ [see \fref{fig1}(a)]. The emitters can be excited by an external field, and we assume that they have a significant optical response in a narrow bandwidth around their resonance frequency $\omega_0$ only. Integrating out the photonic degrees of freedom, and in the Born-Markov approximation, the internal dynamics of the atoms are governed by the master equation $\dot{\rho} = -i[ H ,\rho ] + \mathcal{L}[\rho]$. The Hamiltonian in a frame rotating at the atomic transition frequency $\omega_0$ reads
\begin{equation}
  H = \sum_{ij;i \neq j} \Omega_{ij} \sge_i \seg_j,
  \label{eq:Hamiltonian}
\end{equation}
and the Lindblad operator is
\begin{equation}
  \mathcal{L}[\rho] = \frac{1}{2} \sum_{i,j} \Gamma_{ij} \left( 2\sge_i \rho \seg_j - \seg_i \sge_j \rho - \rho \seg_i \sge_j \right).
  \label{eq:Lindblad}
\end{equation}
The dipole interaction and collective decay matrices with elements $\Omega_{ij}$ and $\Gamma_{ij}$, respectively, are given by
\begin{align}
\Omega_{ij} &= - \frac{3\pi \Gamma_0}{k_0} ~\textrm{Re}\left\{\dbu_i^*\cdot \Gb (\rb_i -\rb_j,\omega_0) \cdot \dbu_j\right\},
\\
\Gamma_{ij} &= \frac{6\pi \Gamma_0}{k_0} ~\textrm{Im}\left\{\dbu_i^* \cdot \Gb (\rb_i -\rb_j, \omega_0) \cdot \dbu_j\right\},
\end{align}
where $\Gb(\rb,\omega_0)$ is Green's tensor in free space, which acts on an oscillating unit dipole according to
\begin{align}
\Gb(\rb,\omega_0) \cdot \dbu &= \frac{e^{i k_0 r}}{4\pi r} \big[ ( \rbu \times \dbu) \times \rbu + 
\\
& + \left( \frac{1}{k_0^2 r^2} -\frac{i}{k_0 r} \right) \left(3 \rbu (\rbu \cdot \dbu) - \dbu \right) \big].
\notag
\end{align}
Here, $\rbu = \rb / |\rb|$ is the position unit-vector, $k_0 = \omega_0 /c$ is the wavenumber associated with the atomic transition, and $\Gamma_0 = \left| \db \right|^2 k_0 ^3 / 3 \pi \hbar \epsilon_0$ is the spontaneous emission rate of a single emitter with dipole moment strength $\db$.

After solving for the atomic density matrix the quantum fields can be obtained from a generalized input-output relation~\cite{asenjo2016,asenjo2017exponential}, which in absence of an external field reads:
\begin{align}
\Eb^+(\rb) = \frac{|\db| k_0^2}{\epsilon_0} \sum_i \Gb(\rb-\rb_i,\omega_0) \cdot \dbu_i  \sge_i.
\label{Eq:Fields}
\end{align}

In the following, we will consider the single-excitation manifold to be significantly occupied only. In this case, for the observables of interest (such as the fields generated by the ring or the excited state population) we can neglect the recycling term (first term in the Lindblad expression). This term accounts for the change in the ground state population, which does not modify said observables. Then, the system can be fully understood from the properties of the eigenstates of an effective Hamiltonian (containing the other two terms of the Lindblad expression only). It reads
\begin{align}
H_{\rm eff} = \sum_{ij}  \left( \Omega_{ij} -i\frac{\Gamma_{ij}}{2} \right) \seg_i \sge_j
\label{Heff}
\end{align}
with $\Omega_{ii} = 0$, as all the emitters are identical and a finite value would represent a global energy shift in the Hamiltonian only.

\section{Collective Excitations and Radiative Properties of a Single Ring}
We start by analyzing a single ring of $N$ dipole-coupled quantum emitters. The eigenstates of $H_{\rm eff}$ define a set of collective modes with associated complex eigenvalues, whose real and imaginary parts correspond to the collective frequency shifts and decay rates, respectively.

For a symmetric ring, where the dipole orientations preserve the rotational symmetry (e.g.\ if the dipoles are oriented perpendicularly to the plane of the ring, or tangentially to the ring), the collective modes in the single-excitation manifold are perfect spin waves given by $\ket{\psi_m} = \tilde{\sigma}^{eg}_m \ket{g}$, with 
\begin{align}
\tilde{\sigma}^{eg}_m = N^{-1/2} \sum_{j=1}^N e^{i m \theta_j} \seg_j.
\end{align}

This is contrary to the case of a finite one-dimensional open chain, where the spin wave approximates the exact solution only~\cite{asenjo2017exponential}. Here, $\theta_j = 2\pi (j-1) /N$ is the angle associated with position $j$ ($j=1,\cdots,N$, see \fref{fig1}), and $m = 0, \pm 1,\pm 2, \cdots, \lceil \pm (N-1)/2 \rceil$ corresponds to the angular momentum of the mode. In these states the single excitation is completely delocalized over all sites, and its angular momentum is well defined. The corresponding eigenvalues are
\begin{align}
\lambda_m = -\frac{3\pi \Gamma_0}{Nk_0} \sum_{j \ell}  e^{i m (\theta_\ell-\theta_j)} G_{j\ell}.
\end{align}
Here, $G_{j\ell} \equiv  \hat{\wp}^*_j \cdot {\Gb} (\rb_j - \rb_\ell,\omega_0) \cdot \hat{\wp}_\ell$ includes the dispersive as well as the dissipative coupling of sites $j$ and $\ell$. Note that, due to the rotational symmetry, the coupling is invariant under a translation along the ring, \emph{i.e.},\ $G_{j\ell} = G_{j+1\ell+1}$. The real and imaginary parts of these eigenvalues define the collective frequency shifts and emission rates of the mode $J_m = \text{Re} \{\lambda_m\}$ and $\Gamma_m = -2 \text{Im} \{\lambda_m\}$, respectively. It is easy to see that the spectrum will be symmetric under $m \leftrightarrow -m$, that is, $\lambda_m = \lambda_{-m}$. We note that the mode $m=0$ is always non-degenerate, whereas the eigenstates with a maximum value of $m$ will be doubly-degenerate if $N$ is odd.

\begin{figure}[t]
\centering
\includegraphics[width=\columnwidth]{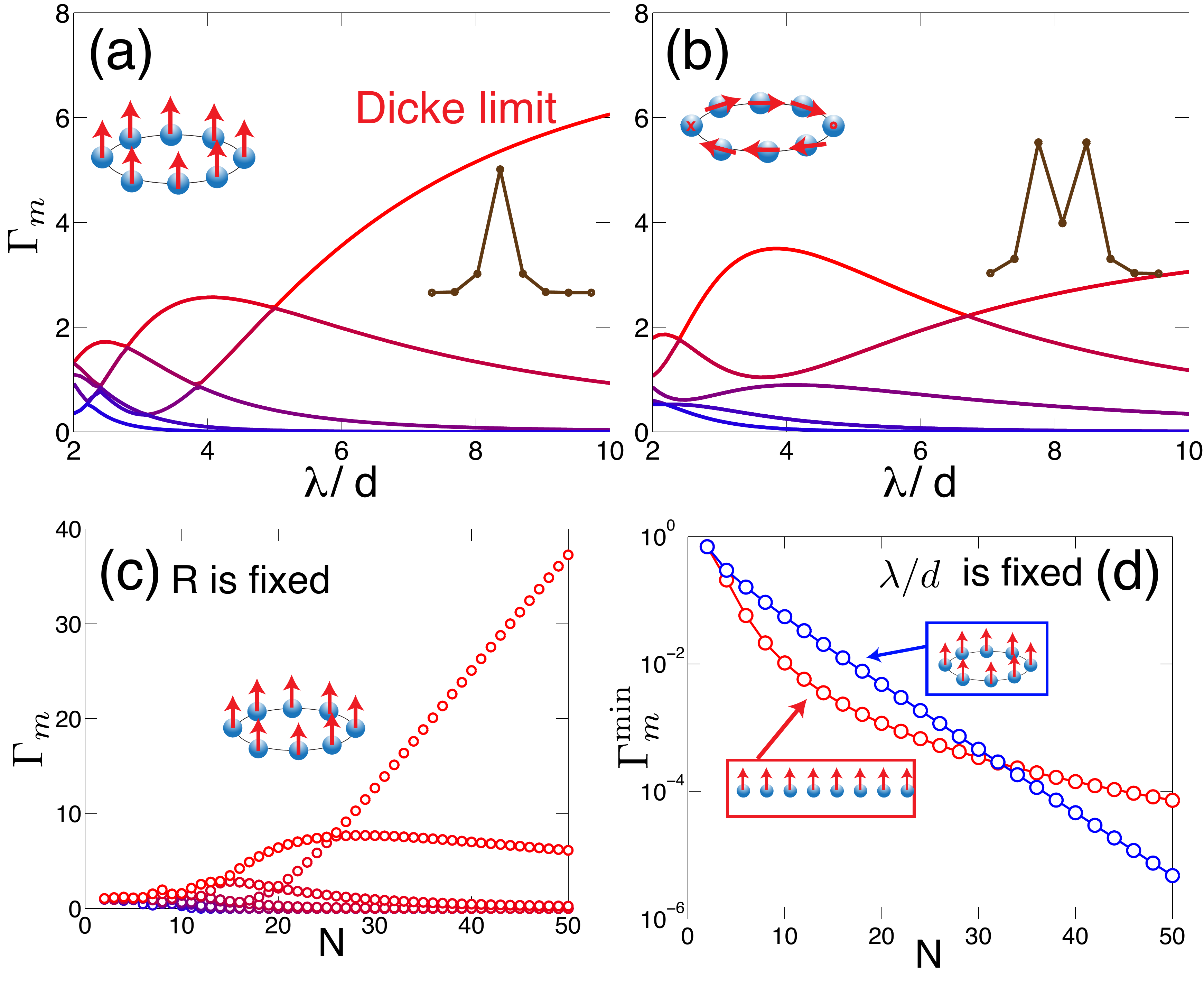}
\caption{\textit{Single ring properties.} {\bf (a)} Collective decay rates $\Gamma_m$ (in units of $\Gamma_0$) as a function of $\lambda/d$, for a ring of $N=8$ emitters with transverse polarization and a single excitation. In the Dicke limit, $\lambda/d \to \infty$, only a single bright mode with a decay rate on the order of $N\Gamma_0$ is present, and $N-1$ modes are dark. {\bf (b)} Identical setup as in (a) but for tangential polarization. Two bright modes arise in the Dicke limit at $m=\pm 1$. {\bf (c)} $\Gamma_m$ (in units of $\Gamma_0$) for a ring of fixed radius with transverse polarization when increasing the density of emitters. For the bright mode, $\Gamma \sim N\Gamma_0$. {\bf (d)} Decay rate (log scale) of the most subradiant eigenmode versus the atom number, for a ring (blue circles) and an open linear chain (red circles), both with $\lambda = 3d$. The lifetime of the most subradiant mode increases exponentially with the atom number.} 
\label{fig2}
\end{figure}

Intuitively, as we decrease the inter-particle distance $d$ with respect to the light wavelength $\lambda = 2\pi / k_0$, we expect to approach the Dicke limit \cite{dicke1954coherence}. In this limit the emitters are so close that the range of the interaction is effectively infinite, yielding a single bright mode decaying at rate $\Gamma = N\Gamma_0$, and $N-1$ perfectly dark modes. This is indeed the case if the polarization is transverse, as it is shown in \fref{fig2}(a), where we have plotted the collective decay rates as a function of the decreasing particle separation. For tangential polarization, on the other hand, there are two bright modes corresponding to $m = \pm 1$ with a decay rate $\Gamma=N\Gamma_0/2$, while $m=0$ is dark by symmetry [see \fref{fig2}(b)].

We can observe this linear scaling of the decay of the most radiant mode with the number of emitters by gradually increasing the density of a ring of constant size. This is shown in \fref{fig2}(c) for transverse polarization. In addition, the covered frequency spectrum becomes increasingly larger as the ring gets denser. The polarization orientation will determine whether the dark or bright modes are lower or higher in energy. For instance, for transverse polarization, bright modes are lower in energy, whereas for tangential polarization (closer to a head-tail configuration of the dipoles) bright modes are higher in energy.

Moreover, the modes of the ring feature extraordinary radiative properties in contrast to an open linear chain. In particular, for a large enough number of emitters there exist subradiant modes whose emission is strongly suppressed. Indeed, if we increase the number of emitters while keeping $\lambda/d$ constant, the system will start to locally resemble an infinite chain. The radiative properties of the infinite chain have been studied in detail in~\cite{asenjo2017exponential}. Interestingly, dark modes in an infinite chain correspond to spin waves characterized by a wavevector along the chain which is larger than $k_0$. In this case the eigenmode generates an evanescent field transversally to the chain and therefore the emitters can guide light perfectly, as if they were a real fiber. For the finite chain, these modes retain a small decay rate since a photon can still scatter off the ends of the chain. However, by bending and closing the chain to form a ring, an increased lifetime of the excitation can be achieved. As observed in Ref.~\cite{asenjo2017exponential}, for a large enough ring, there is an exponential suppression of the decay rate with the number of emitters, in contrast to the polynomial suppression ($\sim N^{-3}$) observed for the open linear chain. A comparison of how the smallest decay rate scales with the atom number in the two cases is shown in~\fref{fig2}(d).

\begin{figure}[t]
\centering
\includegraphics[width=\columnwidth]{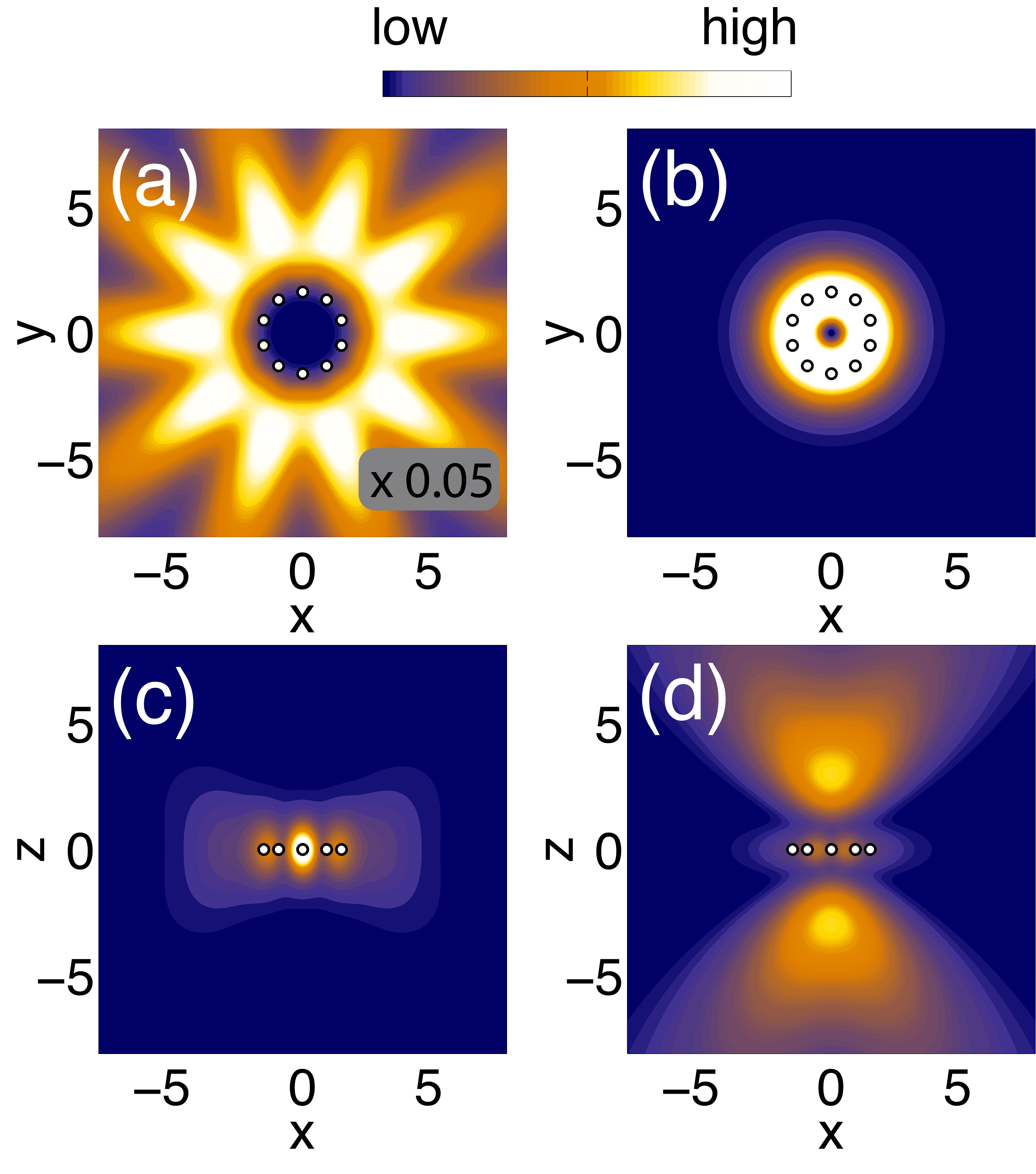}
\caption{\textit{Ring radiation patterns.} The field intensity generated by a single ring with a single excitation and tangential polarization, in the most subradiant mode with $m=\lfloor N/2\rfloor$ [(a), (c)] and in the radiant mode  with $m=0$ [(b), (d)]. The top panels [(a), (b)] show the field in the xy-plane at fixed $z=1.5R$. The bottom panels [(c), (d)] show the field in the xz-plane at fixed $y = 1.5R$. The white circles denote the position of the emitters. The field pattern is remarkably different in the two cases: for the subradiant mode it is evanescent transversally to the ring, and it vanishes at the center of the ring. ($N=10$, $d/\lambda=0.4$).}
\label{fig3}
\end{figure}

Next, we show that the electromagnetic fields generated by a superradiant or a subradiant eigenmode can be radically different. Using Eq.~\eqref{Eq:Fields} we evaluate the fields in \fref{fig3} in real space for a ring with tangential polarization. The fields of the most subradiant (left column) and a radiant eigenmode (right column) are depicted. In addition to the difference in the magnitude of the fields, we find a remarkably distinct radiation pattern. In the case of a subradiant mode, the field is evanescent transversally to the plane that contains the ring, and moreover, it vanishes at the center of the ring. The radiant state, on the other hand, causes strong emission in the transverse direction, and shows an interference maximum at the center of the ring.

\begin{figure}[b]
\centering
\includegraphics[width=\columnwidth]{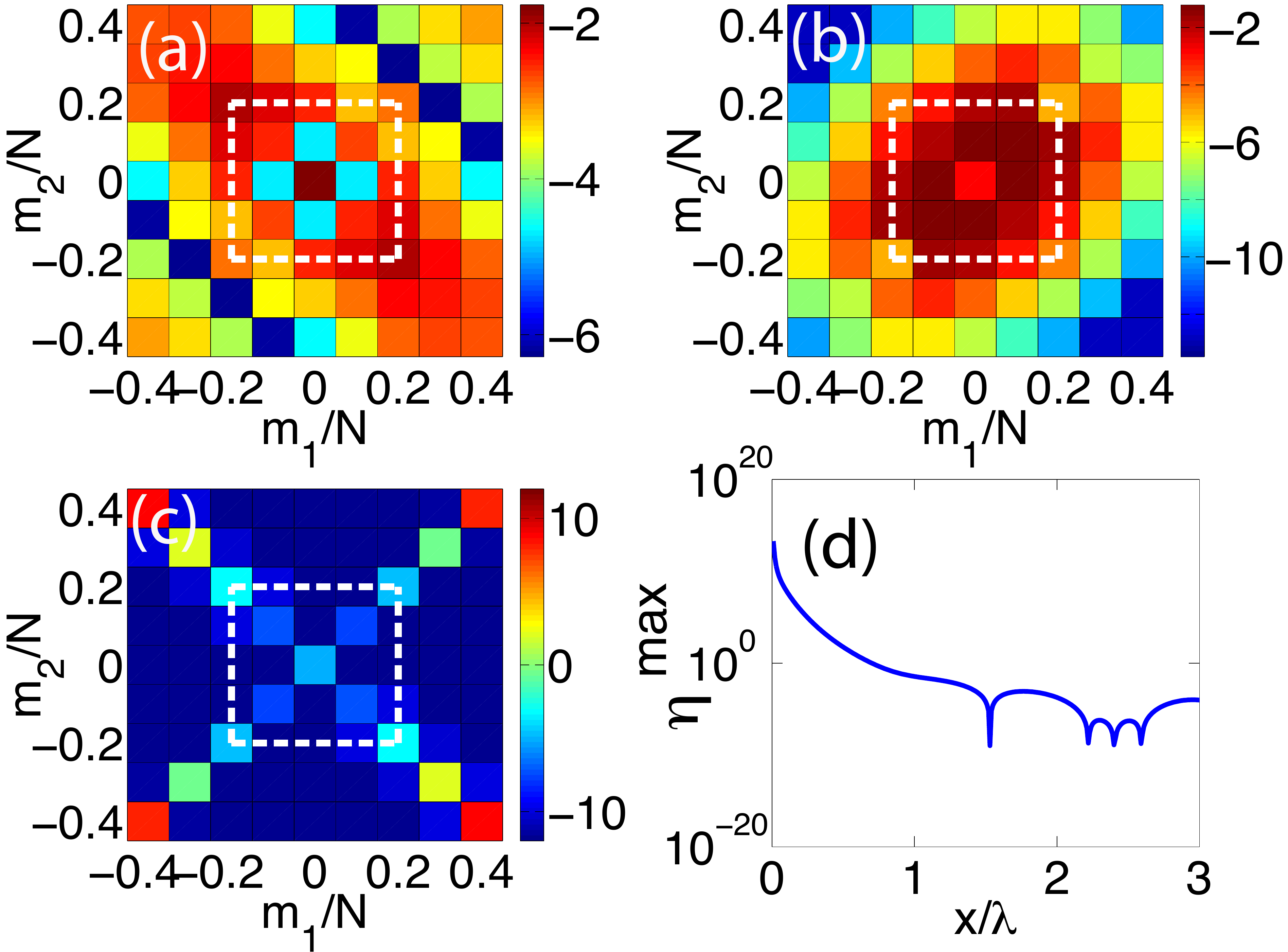}
\caption{\textit{Ring-to-ring coupling.} (a) Absolute values of the dispersive $J_{m_1,m_2}$, (b) dissipative $\Gamma_{m_1,m_2}$  couplings and (c) the ratio $\eta_{m_1,m_2}$ on a logarithmic scale. We consider two rings in spin-wave states with well-defined angular momenta $m_1$ and $m_2$. The rings are contained in the same plane and separated by the distance $x = 0.15 \lambda$, in the configuration shown in \fref{fig1}(b). The dashed white line denotes the light line, beyond which the modes are mainly subradiant. Subradiant states primarily couple dispersively to subradiant states. Moreover, $\eta$ is maximal for subradiant states with an opposite value of angular momentum. (d) The maximum value of $\eta_{m_1,m_2}$ obtained with $m_1=m_2=N/2$, as a function of the ring separation $x$.  ($N=10$, $d/\lambda = 0.1$).}
\label{fig4}
\end{figure}

\section{Tailored Collective Coupling of Two Rings}
As previously discussed in Ref.~\cite{asenjo2017exponential} and presented in more detail above, single rings have very intriguing radiative properties. We will, however, now continue with the study of two coupled rings which lie in the same plane and are separated by a distance $x$ [see \fref{fig1}(b),(c)]. In particular, we study how excitations are transferred from one ring to another with minimal loss. While superradiant states possess the strongest dipole moments and thus couple strongly to neighboring dipoles, they also feature a much faster decay. The effective Hamiltonian, Eq.~\eqref{Heff}, can be rewritten as the sum of intra-ring and ring-ring coupling terms,
\begin{align}
H_{\rm eff} &= \sum_{i, j \in \mathcal{R}_1} \hat{h}_{i,j} +  \sum_{i, j \in \mathcal{R}_2} \hat{h}_{i,j} +  \sum_{\substack{i \in \mathcal{R}_1,\\ j\in \mathcal{R}_2}}  \hat{h}_{i,j}, 
\end{align}
with $\hat{h}_{i,j} = h_{i,j} \seg_i \sge_j$ and $h_{i,j} \equiv -(3\pi\Gamma_0/k_0) {G}_{ij}$. As a shorthand notation, we defined two sets of indices, one for the sites in the first ring, $\mathcal{R}_1=\{1,2,...,N\}$, and one for the sites in the second ring, $\mathcal{R}_2=\{N+1,...,2N\}$, respectively. The last term describes the coupling between the two rings, and it can be written in the angular momentum basis,
\begin{align}
 \sum_{\substack{i \in \mathcal{R}_1,\\ j\in\mathcal{R}_2}}  \hat{h}_{i,j} = \sum_{m_1,m_2} \left( J_{m_1,m_2}-i \frac{\Gamma_{m_1,m_2}}{2} \right) \seg_{m_1,1} \sge_{m_2,2}, \notag
\end{align}
where $J_{m_1,m_2} = \textrm{Re} \{\lambda_{m_1,m_2}\}$ is the dispersive and $\Gamma_{m_1,m_2} = -2 \textrm{Im} \{\lambda_{m_1,m_2}\}$ the dissipative coupling, and
\begin{align}
\lambda_{m_1,m_2} &= \frac{1}{N} \sum_{\substack{i\in\mathcal{R}_1,\\ j\in\mathcal{R}_2}}  h_{i,j} e^{i (m_1 \theta_{i}-m_2 \theta_{j} )}.
\end{align}

In \fref{fig4}(a),(b), we show the dispersive and dissipative couplings as a function of the angular momentum of the two rings $m_1$ and $m_2$. We use the configuration shown in \fref{fig1}(b) with a fixed separation between the two rings $x=\lambda/2$ and tangential polarization. The white dashed line in \fref{fig4}(a)-(c) represents the light line beyond which the states are predominantly subradiant. We observe that subradiant states mainly couple dispersively to other subradiant states, whereas radiant states couple to other radiant states with large dissipation. Furthermore in \fref{fig4}(c) we show the ratio $\eta_{m_1,m_2} \equiv J^2_{m_1,m_2}/ [
4\Delta^2_{m_1,m_2} + \max{\{\Gamma^2_{m_1},\Gamma^2_{m_2}\}}]$, with $\Delta_{m_1,m_2} = |J_{m_1}-J_{m_2}|$. This is a figure of merit that quantifies how efficiently two modes in the two rings are coupled, and thus, how efficiently an excitation can be transferred between them. Remarkably, we find that in the subradiant sector, $\eta$ is non-negligible for states where $m_1 = \pm m_2$ only. Moreover, it is several orders of magnitude larger for $m_1 = - m_2$, that is, for two guided modes that propagate in opposite directions in the two rings. This result indicates that in the subradiant regime the physics is well captured by a two-mode model consisting of the two states with $m$ and $-m$.

We also note that the efficiency in the coupling strongly depends on the particular configuration of the dipoles. As the separation between the rings increases, the maximum value of $\eta$ displays oscillations with an overall decay. This is shown in \fref{fig4}(d), where we have evaluated $\eta^{\rm max} = \eta_{m_1 = N/2, m_2 = N/2}$ for $N=10$ as a function of the rings' separation $x$. Interestingly, a different configuration such as the site-edge arrangement illustrated in \fref{fig1}(c) can yield a dramatically different result. In this case, it is easy to see that due to symmetry, the fields created by the two rings in the $m=N/2$ mode completely destructively interfere yielding a coupling which is exactly zero. 

\section{Efficient Excitation Transfer between Two Rings}
According to the previous results for the couplings, we might expect that if one of the rings is initially prepared in a very subradiant state, the excitation oscillates between the two rings for a very long time before it finally decays.

Up to now we have considered eigenstates of the system which are fully delocalized in space. This gives more insight into the native behavior of the system than investigating single excited sites. However, excitations in reality will often be partially (de-)localized only. This leads to the natural question whether it is still possible to achieve an efficient excitation transport between the two rings for a multi-mode wave packet. In the following, we investigate the dynamics of a wave function initially prepared in the form
\begin{equation}
\ket{\Psi_{i,k}^m} = \frac{1}{\sqrt{n}}\sum_{j\in \mathcal{R}_i} e^{i\theta_j m} e^{-\frac{|\rb_j - \rb_k|^2}{2R^2\Delta\theta^2}}\seg_j\ket{g}. 
\end{equation}

This corresponds to a wave packet with a Gaussian population distribution centered at site $k$ in the $i$th ring with an angular spread $\Delta\theta$ (wave packet width $R \cdot \Delta\theta$) and central momentum $m$. The constant $n$ accounts for the normalization. An infinitely wide wave packet of this form represents the eigenstate given by $m$. On the other hand, if the wave packet is of zero spread, only the atom at the site given by the index $k$ is excited.

For a mode guided by the first ring with momentum $m$, it is only natural that it will invert its direction upon being transported to the second ring. This is a more intuitive picture of the previous result that the coupling is optimal between modes with opposite $m$. Thus, for a finite width wave packet, we expect that it is transferred into a wave packet with the same shape but central momentum $-m$. Therefore, we evaluate the fidelity $\mathcal{F}(t)$ of creating this wave packet in the second ring as
\begin{align}
\mathcal{F}(t) &= \underset{k}{\max}\left\{\braket{\Psi_{2,k}^{-m}|\Psi(t)}\right\},
\end{align}
where we maximize over the site index $k$ in the second ring since we do not know the (center) position of the wave packet created there at all times. The wave function $\ket{\Psi(t)}$ is given by the time evolution with $H_{\rm eff}$ with the initial condition $\ket{\Psi(0)}=\ket{\Psi_{1,k}^m}$.

\begin{figure}[t] 
\centering
\includegraphics[width=\columnwidth]{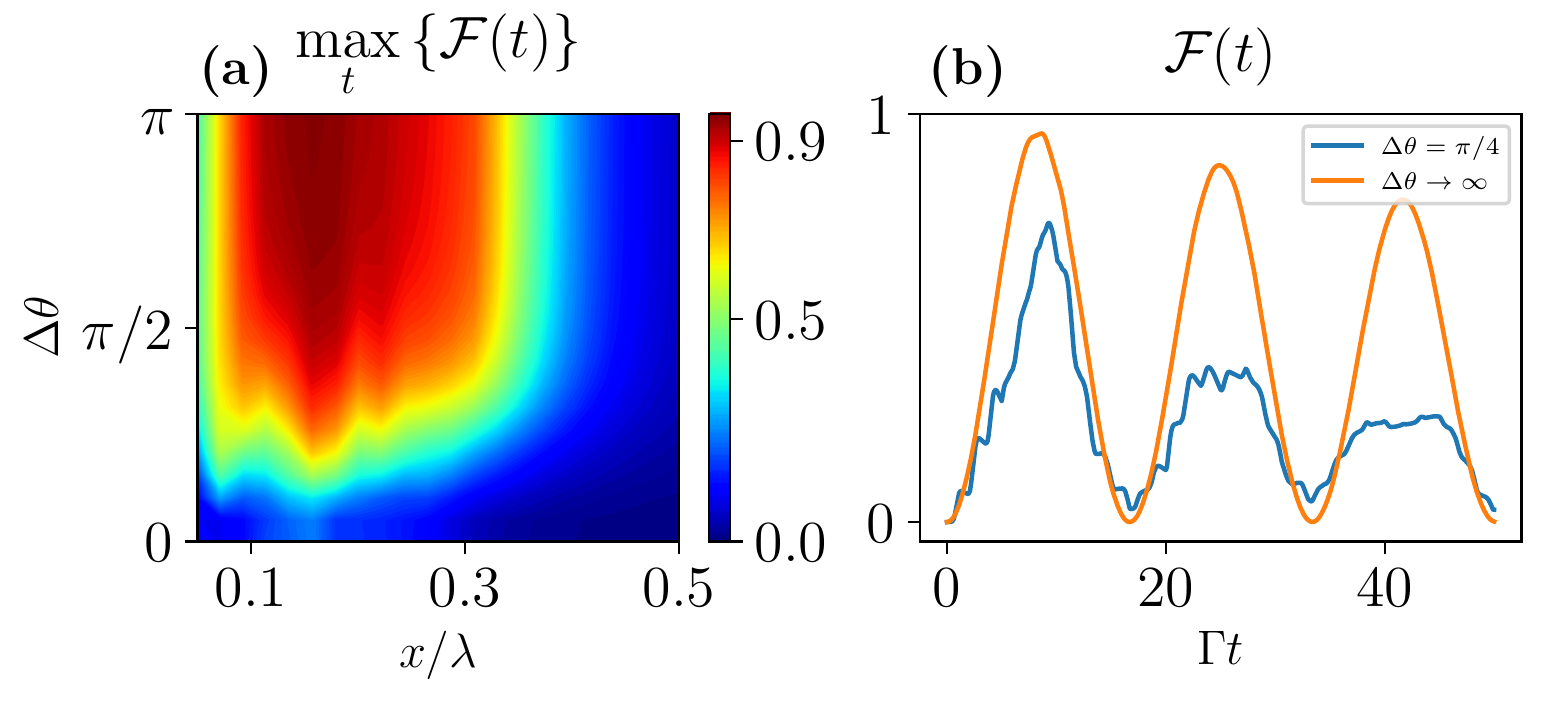}
\caption{\emph{Fidelity of wave packet transport between two rings}. \textbf{(a)} A scan of the maximal fidelity (over time) as a function of the ring separation and the wave packet width for two rings consisting of $N=20$ atoms. For a sufficiently large width and a separation comparable to the inter-particle distance $d=0.1\lambda$, the fidelity is almost unity. \textbf{(b)} The larger the width in real space, the more the wave packet is localized in momentum space thus showing better transport behavior. The separation between the rings was $x=0.15\lambda$. For both (a) and (b) the dipoles were oriented tangentially.}
\label{fig5}
\end{figure}

In \fref{fig5}(a) we show the maximal fidelity during time evolution as a function of the ring separation as well as of the width of the initial wave packet for two rings with $N=20$. We start out with a wave packet centered at the site that is farthest from the second ring. The momentum is chosen to be $m=\lfloor N/4\rfloor$ such that all modes the packet is made up of have momenta of the same sign. As one can see, the fidelity is rather low as long as the width is small, \emph{i.e.,}\ the excitation is localized at one site almost perfectly. However, for a comparably small width in real space already, the wave packet is sufficiently localized in momentum space to exhibit coherent transport. The fidelity grows to values larger than 90\% quickly as the width increases, indicating a reliable transport of a subradiant wave packet from the first ring to the next. Moreover, the transport is best if the separation between the rings is comparable to the inter-particle distance. This is due to the change of the energetic shifts of neighboring atoms with their separation: if the atoms at the points where the rings are closest are too far (or too close) to one another, the shifts vary greatly, effectively detuning these atoms far from the rest. Excitations can then no longer propagate.

In \fref{fig5}(b) we plot the fidelity of the same wave packet being transported as a function of time. The wave packet oscillates between the two rings with the same period for both finite and infinite width. The amplitude, however, damps out considerably faster for the case of a finite width. Nevertheless, a large fidelity is achieved even when the initial state is not a perfect eigenstate of the system.

\section{Conclusions}
We started off by showing that a single ring of dipole-dipole coupled emitters exhibits peculiar radiative properties. Specifically, the lifetime of subradiant states prepared in such a ring increases exponentially with a growing number of atoms. Accordingly, the intensity emitted from a ring when a subradiant state is prepared is largely diminished compared to a superradiant state.

We then showed that if another ring is present, optimal coupling occurs between subradiant states of each ring. An excitation that is sufficiently delocalized and moving with a velocity along one ring that closely corresponds to the momentum of an eigenstate that is subradiant can be transported to another ring with a large fidelity. This reliable coherent transport takes place for a comparably small delocalization already and culminates in damped Rabi-like oscillations between the rings once the excitation is spread over the entire first ring.

Note that beyond the two-level approximation analogous bright and dark states also appear in more complex level structures with several decay channels~\cite{hebenstreit2017subradiance}. Hence much of the physics discussed here should also hold in rings of particles with a more complex internal structure. 

\begin{acknowledgments}
\section{Acknowledgements}
We acknowledge financial support by the Austrian Science Fund (FWF) through projects P29318-N27 (H.~R. and L.~O.) and the DK-ALM W1259-N27 (D.~P.) as well as from ERC Starting Grant FOQAL, MINECO Plan Nacional Grant CANS, MINECO Severo Ochoa Grant No. SEV 2015-0522, CERCA Programme/Generalitat de Catalunya, AGAUR Grant 2017 SGR 1334, and Fundacio Privada Cellex (D.C.). We thank N. van Hulst and  C. Genes for helpful discussions and comments. Part of the numerical simulations were performed with the open-source framework QuantumOptics.jl~\cite{kramer2018quantumoptics}. After completion we became aware of related work on coupling between two planar arrays using collective dark states~\cite{guimond2019subradiant}.
\end{acknowledgments}

\bibliography{darkring}
\end{document}